\documentclass{jetpl}
\twocolumn

\lat

\title{Spin-orbit lateral superlattices: energy bands and spin polarization in 2DEG}

\rtitle{Spin-orbit lateral superlattices: energy bands and spin
polarization in 2DEG}

\sodtitle{Spin-orbit lateral superlattices: energy bands and spin
polarization in 2DEG}

\author{V.\,Ya.\,Demikhovskii,
D.\,V.\,Khomitsky\/\thanks{e-mail: khomitsky@phys.unn.ru}}

\rauthor{V.\,Ya.\,Demikhovskii, D.\,V.\,Khomitsky}

\sodauthor{Demikhovskii, Khomitsky}

\address{Department of Physics, University of Nizhny Novgorod, Gagarin Avenue 23,
603950 Nizhny Novgorod, Russia}

\dates{14 March 2006}{*}

\abstract{The Bloch spinors, energy spectrum and spin density in
energy bands are studied for the two-dimensional electron gas (2DEG) with
Rashba spin-orbit (SO) interaction subject to one-dimensional (1D) periodic
electrostatic potential of a lateral superlattice. The space symmetry
of the Bloch spinors with spin parity is studied. It is shown that
the Bloch spinors at fixed quasimomentum describe the standing
spin waves with the wavelength equal to the superlattice period.
The spin projections in these states have the components both parallel and
transverse to the 2DEG plane. The anticrossing of the energy dispersion curves
due to the interplay between the SO and periodic terms is observed,
leading to the spin flip. The relation between the spin parity and
the interband optical selection rules is discussed, and the effect of
magnetization of the SO superlattice in the presence of external
electric field is predicted.}


\begin{document}

\maketitle

\section*{Introduction}

In the past years, an increasing attention has been drawn to the spin related
phenomena in semiconductor structures. This research area has developed in
the new branch of condensed matter physics and spin electronics.
The problem of spin-dependent quantum states and transport phenomena in these
systems are currently attracting a lot of interest also due to their potential
for future electronic device applications.

In two-dimensional semiconductor heterostructures the spin-orbit
interaction is usually dominated by the Rashba coupling
\cite{Rash60} stemming from the structure inversion asymmetry of
confining potential. The low-dimensional semiconductor structures
with SO interaction were studied theoretically in numerous papers
[2 \ch 5], including the 1D periodic systems with SO coupling
\cite{periodic},\cite{Bryksin}.

The effects of spin splitting in 2DEG were investigated
experimentally by the magnetotransport studies, in particular, by
Shubnikov - de Haas oscillations \cite{experiment}. The role of
other SO terms, such as Dresselhaus term, can be estimated, for
example, by optical methods \cite{Ganichev04}. As it was
demonstrated by the experiments, in the SO structures interesting
effects may be observed, such as the spin Hall effect \cite{Hall}
and the spin-galvanic effect \cite{Ganichev03}.

In the present paper we study quantum states and the electron spin
distribution in a system combining the spin-splitting phenomena
caused by the SO interaction and the external gate-controlled
periodic electric potential. We thus want to investigate the spin
orientation and spin polarization that can be achieved in
currently manufactured gated semiconductor structures with lateral
surface superlattice. For example, the 1D superlattice can be
fabricated by the metal gate evaporation with typical period of
$50-200$ {\it nm}. We use the value of lateral period in the
$x$-direction to be $a=60$ {\it nm} which gives us the energy scale
$\pi^2 \hbar^2/2ma^2$ of the order of $2$ {\it meV} for the effective
mass $m^*=0.067m_0$ in GaAs. The values of Rashba coupling constant
for the most important semiconductors are in the range of
$(1 \ldots 5) \cdot 10^{-11}$ {\it eVm}. It is known also that the Rashba
coupling strength can be modified by the gate field by up to 50\% \cite{Miller03}.
Below in our calculations we use the value $\alpha=5 \cdot 10^{-11}$ {\it eVm}
which gives the typical shift of the parabolic dispersion curves $k_{SO}$
to be of the order of $\pi/a$. So, in the structure studied in the manuscript
the electron kinetic energy $\pi ^2\hbar^2k^2/2m$ will be comparable to
the Rashba energy $\alpha k$ which makes the effects of SO interaction
and periodic potential distinguishable. It should be mentioned
also that the energy scale studied in our paper means that the
effects discussed in the manuscript can be clearly observed
experimentally at helium temperatures.

\section {Quantum states}

The Hamiltonian of our problem
\begin{equation}
\hat{H}=\hat{H}_0+V(x)
\label{ham}
\end{equation}

\noindent is the sum of the Rashba Hamiltonian ${\hat H}_0$ with the SO coupling
strength $\alpha$,
\begin{equation}
\hat{H}_0=\frac{\hat{p}^2}{2m^*}+ \alpha
(\hat{\sigma}_x\hat{p}_y-\hat{\sigma}_y\hat{p}_x),
\label{h0}
\end{equation}

\noindent and the one-dimensional periodic potential $V(x)$ of
a 1D superlattice with the period $a$. We choose the simplest form of
the periodic potential
\begin{equation}
V(x)=V_0 \cos \frac{2\pi x}{a},\label{vx}
\end{equation}

\noindent where the sign and the magnitude of $V_0$ can be controlled,
for example, by an external gate, leaving us a wide interval
of possible $V_0$ amplitudes with an order of several {\it meV}.

\subsection{Perturbation approach}

\begin{figure}[t]
  \centering
  \includegraphics[width=85mm]{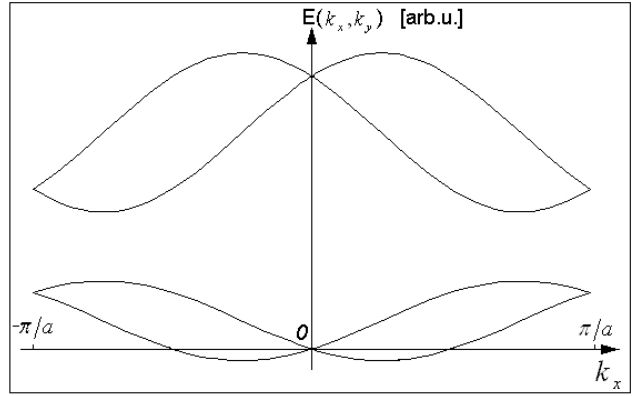}
  \caption{{\bf Fig.\ref{fepert}} \quad Energy bands at $k_y=0$ double-split by
            the SO perturbation. The degeneracy is not lifted at
            $k_x=0, \, \pm \pi/a$, \, $k_y=0$ since the derivative
            $\frac{\partial \varepsilon_m}{\partial p_x}$ of the unperturbed band
            dispersion vanishes at these points.}
\label{fepert}
\end{figure}

Qualitatively the formation of the SO-split bands can be seen in
the perturbation approach applied in the problem of quantum states
in quantum wires \cite{Governale} and in the tight-binding
approximation for the SO superlattices \cite{Bryksin}. Namely, at
$k_y=0$ the Hamiltonian (\ref{ham}) is written as
$\frac{\hat{p}^2}{2m^*}+V(x)- \alpha \hat{\sigma}_y\hat{p}_x$.
Considering the SO terms $- \alpha \hat{\sigma}_y\hat{p}_x$ as a
perturbation, one can choose the zero-order wavefunction as
\begin{equation}
\Psi(x)=\psi_{mk}(x)
\left(
\begin{array}{c}
1 \\
\pm i
\end{array}
\right)
\label{zrwf}
\end{equation}

\noindent where $\psi_{mk}(x)$ are the eigenstates of the
Hamiltonian $\frac{\hat{p}^2}{2m^*}+V(x)$ and $m$ is the band number.
Thus, in the first order of the perturbation theory the energy spectrum
will be determined by the expression
\begin{equation}
E_m(k_x,k_y=0)=\varepsilon_m(k_x) \pm \frac{\hbar k_{SO}}{m^*}
\frac{\partial \varepsilon_m}{\partial p_x} \label{enpert}
\end{equation}

\noindent where $\varepsilon_m(k_x)$ is the band spectrum in the
1D periodic potential $V(x)$ without the SO interaction.
Here we have used the relation
$\frac{1}{m^*}\langle \psi_{mk}\mid {\hat p}_x \mid
\psi_{mk}\rangle= \frac{\partial \varepsilon_m}{\partial p_x}$ for
derivation of Eq.\ref{enpert}. An example of the energy band
spectrum of two lowest bands at $k_y=0$ each double-split by the SO
interaction is given in Fig.\ref{fepert}. One can see that since
the derivative $\frac{\partial \varepsilon_m}{\partial p_x}$
vanishes at the border of the Brillouin zone (BZ) $k_x=\pm \pi/a$, \,
$k_y=0$ and at the BZ center, the degeneracy is not lifted in these
points, being lifted at finite $k_y$ only.

\subsection{Bloch spinors}

\noindent In the presence of both SO coupling and periodic potential we
construct the two-component eigenstate of the Hamiltonian (\ref{ham})
\begin{equation}
\psi_{\bf k}(x,y)=e^{ik_y y}
\left(
\begin{array}{c}
\psi^1_{\bf k}(x) \\
\psi^2_{\bf k}(x)
\end{array}
\right)
\label{psi12}
\end{equation}

\noindent as a superposition of two-component spinors which are the eigenstates of
the Rashba Hamiltonian (\ref{h0}). The wavevectors of the basis states in this superposition
are shifted by the reciprocal lattice vector ${\bf b}$ of the supelattice:
\begin{equation}
{\bf k}_n={\bf k}+n{\bf b}=\left(k_x+\frac{2\pi}{a}n, \quad k_y\right),
\label{kn}
\end{equation}

\noindent $n=0,\pm1,\pm2,\ldots$. The eigenstate in  band $m$ thus has the form

\begin{equation}
\psi_{m\bf k}=\sum_{\lambda n} a^m_{\lambda n} ({\bf k})
\frac{e^{i{\bf k}_n{\bf r}}}{\sqrt{2}} \left(
\begin{array}{c}
1 \\
\lambda e^{i\theta_n}
\end{array}
\right),
\quad \lambda = \pm 1
\label{wf}
\end{equation}

\noindent where ${\bf k}$ is the quasimomentum in the 1D Brillouin
zone and $\theta_n={\rm arg}[k_y - ik_{nx}]$. After substituting the wavefunction
(\ref{wf}) into the Schr\"odinger equation the coefficients $a^m_{\lambda n}$ are
determined by the standard eigenvalue problem
\begin{equation}
\sum_{\lambda'n'} \left[ \left(E^{R}_{n'\lambda'}-E\right)\delta_{\lambda n \lambda' n'}
 +V^{\lambda \lambda'}_{nn'} \right] a^m_{\lambda' n'}=0, \label{sys}
\end{equation}

\noindent where $E^{R}_{\lambda n}$ is the energy of a free Rashba quantum state
\begin{equation}
E^R_{\lambda}=\hbar^2 k^2/2m + \lambda \alpha k
\label{er}
\end{equation}

\noindent taken at the point in ${\bf k}$-space defined by
(\ref{kn}), i.e. $E^{R}_{n \lambda}=E^{R}_{\lambda}(k_x+2\pi n /a,
k_y)$. The matrix elements in the system (\ref{sys}) have the form
\begin{equation}
\begin{array}{c}
V^{\lambda \lambda'}_{nn'}= V_0A_{nn'}(1+\lambda \lambda'
e^{i(\theta_n-\theta_{n'})}), \\
A_{nn'}=\frac{1}{2}\delta_{n,n'\pm 1}, \quad n=n'\pm 1.
\end{array}
\label{vcos}
\end{equation}

\noindent The structure of matrix elements (\ref{vcos}) determines
the classification of energy bands and gaps in the SO superlattice.
The dependence of the matrix elements (\ref{vcos}) on the quantum
numbers $k_x$ and $k_y$ can be obtained directly from Eq.(\ref{vcos}).
The matrix elements $V^{+-}_{n,n\pm 1}$ and $V^{-+}_{n,n\pm 1}$ describe
the coupling between the nearest-neighboring states (\ref{kn})
with $n'=n \pm 1$ and with the opposite indices $\lambda \ne
\lambda'$ labeling different up- and down- Rashba bands
(\ref{er}). Here the superscript $\pm$ labels the index $\lambda =
\pm 1$. We shall see below that these elements produce the energy
gaps located inside the Brillouin zone. The elements
$V^{++}_{n,n\pm 1}$ and $V^{--}_{n,n\pm 1}$ are responsible for
the coupling between the states of the same Rashba index
$\lambda=\lambda'$. Such elements could open gaps on the borders
of the Brillouin zone $k_x=\pm \pi/a$. However, one can see from
(\ref{vcos}) that they vanish for the case of pure electrostatic
periodic potential. Then, for the values of at $k_y \gg \pi/a$
the elements with opposite indices $\lambda \ne \lambda'$,
i.e. $V^{+-}_{n,n\pm 1}$ and $V^{-+}_{n,n\pm 1}$ decrease to zero
while the elements with $\lambda = \lambda'$ approach their
maximum values.

\subsection{Energy spectrum}

\begin{figure}[t]
  \centering
  \includegraphics[width=85mm]{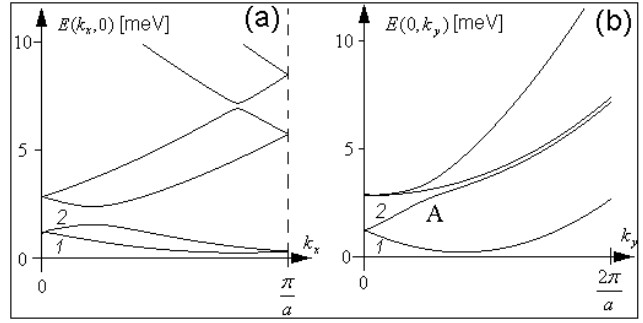}
  \caption{{\bf Fig.\ref{fen}} \quad Energy spectrum at $\alpha=5 \cdot 10^{-11}$ eVm,
  $V_0=1.7$ meV shown (a) as a function of $k_x$ at $k_y=0$ and (b) as a function of $k_y$
  at $k_x=0$. In the latter case the anticrossing takes place at the point {\it A} for band 2.}
\label{fen}
\end{figure}

The energy spectrum calculated with the help of system (\ref{sys})
is shown in Fig.\ref{fen}. Here in Fig.\ref{fen}a we give an
example of energy spectrum $E(k_x)$ at fixed $k_y=0$ and in
Fig.\ref{fen}b the $k_y$-dependence of the same spectrum for
$k_x=0$. In accordance with the Kramers theorem the symmetry
$E_{m \uparrow}(\bf k)=E_{m \downarrow}(\bf -k)$ together with the
symmetry $k_{x,y} \to -k_{x,y}$ takes place, and we thus show the
spectrum only at positive $k_x$ and $k_y$. One can see that the
spin degeneracy at $k_y=0$ is not lifted at the center and at the
borders of the BZ $k_x=\pm \pi/a$. This result was also obtained
earlier in the perturbation approach, see Fig.\ref{fepert}. The
nature of this effect is due to the specific $k_x$ and $k_y$
dependence of matrix elements (\ref{vcos}). The elements
$V^{\lambda=\lambda'}_{nn\pm1}$ which are responsible for the
degeneracy lifting at $k_x=\pm \pi/a$ and $k_y=0$ vanish at
$k_y=0$.  The other set of elements
$V^{\lambda=-\lambda'}_{nn\pm1}$ is non-zero at $k_y=0$, and it
opens the gaps inside the BZ. The $k_y$-dependence of the energy
bands at $k_x=0$ is shown in Fig.\ref{fen}b. The degeneracy at
$k_y=0$ is lifted at finite $k_y$ by mutual influence of linear
$k_y$ terms in (\ref{h0}) and by the matrix elements (\ref{vcos}).
At certain conditions the anticrossing of the dispersion curves
from different bands \cite{Governale} may take place. An example
of the anticrossing effect can be see in Fig.\ref{fen}b near the
point {\it A} for the band 2. Below we shall see that the anticrossing
leads to the spin flip in the $(k_x,k_y)$ plane for quantum states near
the anticrossing point.

\section{Spin polarization}

The control on the spin polarization is crucial for practical
implementation of spintronics. Below we show that in the SO
superlattice the standing spin wave with the period equal to the
superlattice period is formed. We discuss the space distribution
of the spin density for the states with different quasimomenta
${\bf k}$ and the distribution of spin expectation values in the
Brillouin zone. The latter describes the mean value of spin
polarization for the electrons travelling in different directions.
We have calculated the spin density
\begin{equation}
S_{i{\bf k}}(x,y)=(\psi_{\bf k})^{\dagger } {\hat \sigma}_i \psi_{\bf k}
\label{sdens}
\end{equation}

\noindent for a quantum state $\psi_{\bf k}$ in a given band and
after the space integration we obtained the vector field of 2D spin
expectation values $(\sigma_x({\bf k}), \sigma_y({\bf k}))$ in
the Brillouin zone:
\begin{equation}
\sigma_i({\bf k})=\langle \psi_{\bf k} \mid {\hat \sigma}_i \mid
\psi_{\bf k} \rangle. \label{spins}
\end{equation}

\begin{figure}[t]
  \centering
  \includegraphics[width=85mm]{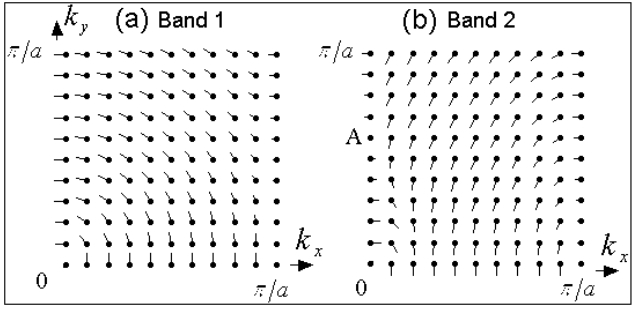}
  \caption{{\bf Fig.\ref{fsk}} \quad Spin polarization shown in one quarter of the BZ for of
          the lowest band 1 (a) and the next band 2 (b). In the latter case the anticrossing
          takes place at point {\it A}, leading to the spin flip in the $(k_x,k_y)$ plane.
          The parameters are the same as in Fig.\ref{fen}.}
\label{fsk}
\end{figure}

\noindent In Fig.\ref{fsk} we show the calculated distribution of
$(\sigma_x, \sigma_y)$ for two lowest bands 1 and 2 shown in
Fig.\ref{fen}. One can see that the spin polarization is
qualitatively modified by periodic potential. The uniform curl distribution
of spins which typical for the 2DEG with SO interaction and without
the periodic potential is conserved only near the BZ center.
It can be seen in Fig. \ref{fsk}a that the curl distribution is destroyed at
the borders $k_x=\pm \pi/a$ of the BZ. The principal difference is that
at $k_x=\pm \pi/a$ the spins are polarized along $x$ axis and $\sigma_y=0$,
and a new type of singularity appears at $k_x=\pm \pi/a$, $k_y=0$.
More complicated picture shown in Fig.\ref{fsk}b takes place for
the spin polarization in the next energy band 2 shown in Fig.\ref{fen}.
We see that at the BZ center the curl topology of spin polarization
(with reversed angular velocity since $\lambda=1$ for the upper Rashba band)
is again unchanged. The greatest changes from the uniform curl distribution
also happen near the borders $k_x=\pm \pi/a$ of the BZ where a new curl
has emerged. Another important feature of this spin distribution
is the spin flip at the point {\it A} for band 2 shown on the axis $k_x=0$ in
the BZ. One can easily establish this point as the anticrossing
point for the energy spectrum $\varepsilon=\varepsilon(k_x=0,k_y)$
shown in Fig.\ref{fen}b. It should be mentioned that such effect
was studied previously in quantum wires with SO interaction
\cite{Governale}. Besides, one can mention in Fig.\ref{fsk} that
at certain points in the BZ the spin average values are decreased.
The nature of such behavior is the mutual influence of SO terms
and periodic potential and will be discussed below.

Another effect which can be promising for the application of SO
superlattices can be found when an external electric field in the
$y$ direction is applied. Under these conditions the spin
distribution in the BZ is shifted homogeneous along $k_x$ which
can produce the magnetization along $x$ direction. It is clear
that the sign of this magnetization changes with the direction of
the $E_y$ component of the external electric field and thus can be
easily controlled.

\section{Spinor symmetry and selection rules}

\subsection{Spin parity}

The symmetry of the confinement potential $V(x)=V(-x)$ leads to
the existence of an additional quantum number called the spin
parity \cite{Debald}. Namely, the Hamiltonian (\ref{ham}) commutes
with the spin parity operator
\begin{equation}
{\hat S}_x={\hat P}_x {\hat \sigma}_x
\label{sp}
\end{equation}

\noindent where ${\hat P}_x$ is the inversion operator of the $x$
coordinate, ${\hat P}_x f(x)=f(-x)$. We shall study the
$x$-dependence of the spinor components $\psi^{1,2}(x)$ of
quantum state (\ref{psi12}). Below we demonstrate that the Bloch spinors in our problem
taken at the points ${\bf k}=(k_x=0,\pm \pi/a, \, k_y)$ can be labeled by
a certain quantum number $s$ which is the spin parity. In this case
the spinor components $\psi^{1,2}(x)$ of (\ref{psi12}) satisfy the following symmetry
relations:
\begin{equation}
\psi^{1,2}_{\bf k}(x)=s \, \psi^{2,1}_{\bf k}(-x), \label{psis}
\end{equation}

\noindent where $s=\pm 1$ is the spin parity.

\begin{figure}[ht]
  \centering
  \includegraphics[width=80mm]{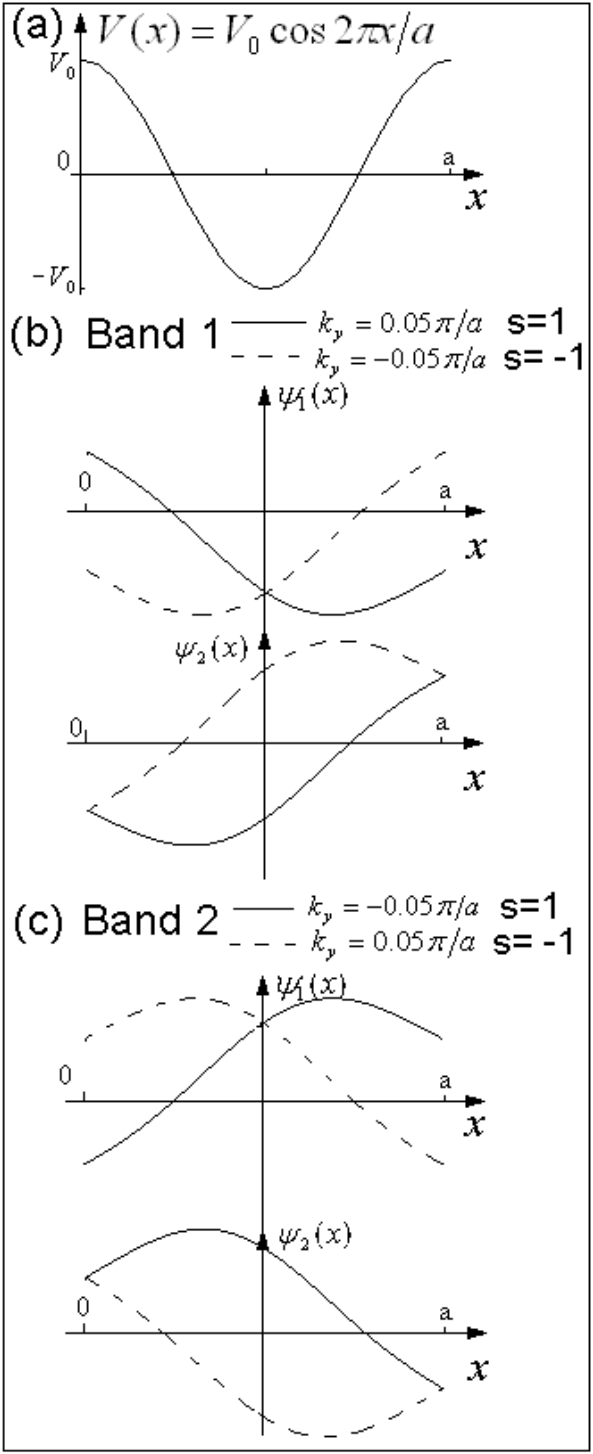}
  \caption{{\bf Fig.\ref{fpsis}} \quad (a) supelattice potential V(x) and
            (b), (c) space dependence of the spinor components $\psi^{1,2}(x)$
            in a superlattice elementary cell at $k_y=0.05 \pm \pi/a$
            shown (b) for the lowest band 1 and (c) for the next band 2.
            The spin parity $s$ changes from band 1 to band 2 and when $k_y \to -k_y$.}
\label{fpsis}
\end{figure}

In Fig.\ref{fpsis}b,c we show the $x$-dependence of the spinor
components $\psi^{1,2}(x)$ in a superlattice elementary cell
together with the superlattice potential $V(x)$ shown in
Fig.\ref{fpsis}a. One can see that the spinor components
$\psi^{1,2}(x)$ are real at $k_x=\pm \pi/a$. The states are taken
at opposite $k_y$ momentum components and for two neighboring
energy bands 1 and 2. By comparing Fig.\ref{fpsis}(a) and
\ref{fpsis}(b), \ref{fpsis}(c) one can see that the space symmetry
of a particular wavefunction on the superlattice period does not
follow the symmetry of the superlattice potential $V(x)$. The reason
is the SO coupling which produces the space shift of the components
$\psi^{1,2}(x)$ of the spinor (\ref{psi12}). Such effect was observed
previously for the case of quasi one-dimensional SO systems \cite{Moroz}.
Another important feature which can be seen in Fig.\ref{fpsis} is that
the spin parity changes both from the lowest band 1 to the higher band 2
and it changes also under the reflection $k_y \to -k_y$. In the next Section
we shall see how these properties influence on the optical selection rules
for the interband transitions.

The symmetry relation (\ref{psis}) leads to the following property of the
spin density distribution (\ref{sdens}):
\begin{equation}
S_{y,z}(x)=-S_{y,z}(-x).
\label{sdsymm}
\end{equation}

\begin{figure}[t]
  \centering
  \includegraphics[width=85mm]{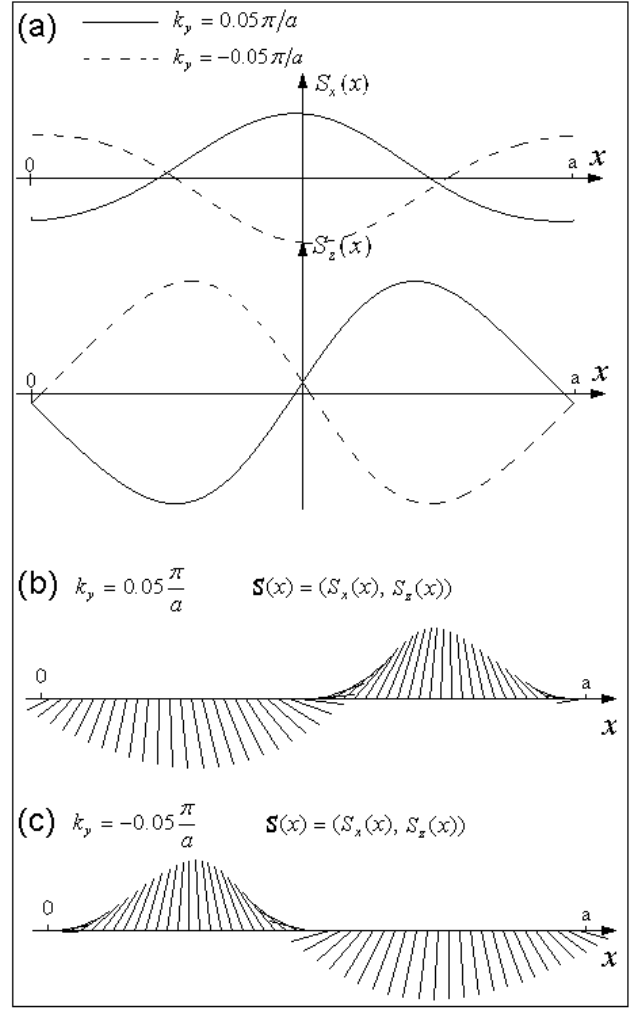}
  \caption{{\bf Fig.\ref{fspx}} \quad (a) spin density distribution $S_x(x)$, \, $S_z(x)$ and (b), (c)
  the $x$-dependence of the two-dimensional spin vector $(S_x(x), \, S_z(x))$ shown for the lowest band 1
  (see Fig.\ref{fen}) at $k_x=\pi/a$, \, $k_y=\pm 0.05 \pi/a$. The spin components switch their signs when
  the $k_y$ momentum projection is reflected due to the change of the spin parity quantum number.
  Inside the superlattice elementary cell one can observe in (b) and (c) the spin standing wave with
  the spin wriggling in opposite directions for the opposite $k_y$ components of the electron momentum.}
\label{fspx}
\end{figure}

\noindent Since the spinor components $\psi^{1,2}(x)$ are real at
$k_x=\pm \pi/a$, the $S_y$ component of spin density vanish
identically. So, we calculate the space distribution of the other
components, $S_x(x)$ and $S_z(x)$ taken for the point $k_x=\pi/a$,
\, $k_y=\pm 0.05 \pi/a$ in the Brillouin zone of the lowest energy
band (the corresponding spin polarization which is the space
integral of $S_{x,y}$ is shown above in Fig.\ref{fsk}(a)). It
should be stressed that in SO superlattice the $S_z$ component
transverse to the superlattice plane appears while it is absent in
pure Rashba 2DEG. The results for $S_x(x)$, \, $S_z(x)$, and the
$x$-dependence of the two-dimensional vector $(S_x(x), \, S_z(x))$
are shown in Fig.\ref{fspx}. One can see that the average values
of the spin density across the superlattice cell shown in
Fig.\ref{fspx} are close to zero which can be seen also in
Fig.\ref{fsk} in the vicinity of the point $k_x=\pi/a$, \,
$k_y=\pm 0.05 \pi/a$ where no clear indication of an arrow is
seen. In Fig.\ref{fspx} it is clear also that the spin components
change their signs when the $k_y$ momentum projection is changed,
thanks to the change of the spin parity quantum number observed
above. In more details this effect can be explained with the help
of the operator composition for the reflection in the reciprocal
space. Namely, the transformation ${\hat K}_y=k_y \to -k_y$ at the
Brillouin zone boundary $k_x=\pm \pi/a$ is equivalent to the
inversion operator ${\hat K}={\bf k} \to -{\bf k}$ since the lines
$k_x=\pm \pi/a$ are topologically identical and one can write
${\hat K}_y{\hat K}_x = {\hat K}$ where ${\hat K}_x=k_x \to -k_x$
and ${\hat K}_x={\hat 1}$ at the Brillouin zone border $k_x=\pm
\pi/a$. Thus, the reflection of $k_y$ here produces the same
effect as the inversion ${\bf k} \to -{\bf k}$, the latter giving
us the spin flip. As it was mentioned earlier, the spin parity and
the corresponding symmetry relation for it at $k_y \to -k_y$
should exist also at $k_x=0$. Indeed, at $k_x=0$ the operation
$k_y \to -k_y$ is also identical to ${\bf k} \to -{\bf k}$ thus
producing the change $s \to -s$ and making a spin flip.

By looking onto Fig.\ref{fspx}(b,c), one can call the spin density
distribution inside the superlattice elementary cell as a standing
spin wave. From Fig.\ref{fspx}(b,c) one can see that the wriggling
takes place with the opposite direction for the opposite $k_y$
components of the electron momenta in accordance with the symmetry
properties discussed above.

\subsection{Selection rules}

The analysis of the optical properties of the SO superlattices
requires the knowledge of corresponding selection rules. We
discuss here the selection rules for the operator of the dipole
momentum ${\hat x}$. The symmetry analysis of the wavefunction
carried out above will help us to define the selection rules for
the certain points in the BZ $k_x=0, \pm \pi/a$ which are often
correspond to the energy minima and maxima. By considering the
symmetry properties (\ref{psis}) of the spinor (\ref{psi12})
one can see that the matrix element

\begin{equation}
M^{ss'}_{mm'k}=\langle \psi_{msk} \mid {\hat x} \mid \psi_{m's'k} \rangle
\label{xdip}
\end{equation}

\noindent for the direct transitions between the bands $m$ and
$m'$ having the spin parity $s$ and $s'$ is equal to

\begin{equation}
M^{ss'}_{mm'k}=(1-ss')M_{mm'k}. \label{xint}
\end{equation}

\noindent Here $M_{mm'k}$ is an integral independent of spin
parity and in general it is non-zero. So, the direct optical
transitions are allowed only between the states of the opposite
spin parity. For example, for our problem the states in two lowest
neighboring spin-split bands 1 and 2 discussed above satisfy to this
condition and the direct transitions between them are thus
possible.

In conclusion we would like to mention also that in the presence of
and external constant electric field parallel to the superlattice direction
one can expect a simultaneous appearance of wriggling electron trajectories
({\it zitterbewegung} \cite{zitt}) and Bloch oscillations. Such situation may lead
to non-trivial dynamics and transport of charged particles, and we plan to investigate
these problems in a separate paper.

\section*{Acknowledgements}

This work was supported by the Program "Development of the
Higher school research potential" (2006), by the Russian Foundation for
Basic Research Grant No.05-02-16449, by the CRDF Award
RUX0-001-NN-06, and by the Dynasty Foundation.


\begin{thebibliography}{20}

\bibitem{Rash60} E.I. Rashba, Fiz. Tverd. Tela (Leningrad) {\bf 2}, 1224
(1960) [Sov. Phys. Solid State {\bf 2}, 1109 (1960)];
Y.A. Bychkov and E.I. Rashba, Pis'ma Zh. \'Eksp. Theor. Fiz. {\bf 39}, 66 (1984)
[JETP Lett. {\bf 39}, 78 (1984)].

\bibitem{Moroz}
A.V. Moroz and C.H.W. Barnes, Phys. Rev. B {\bf 60}, 14272 (1999);

\bibitem{theory} F. Mireles and G. Kirczenow, Phys. Rev. B {\bf 64}, 024426 (2001);
X.F. Wang and P. Vasilopoulos, Phys. Rev. B {\bf 67}, 085313 (2003).

\bibitem{Governale}
M. Governale and U. Z\"ulicke, Phys. Rev. B {\bf 66}, 073311 (2002).

\bibitem{Debald} S. Debald and B. Kramer, Phys. Rev. B {\bf 71}, 115322 (2005).

\bibitem{periodic} X.F. Wang, Phys. Rev. B {\bf 69}, 035302 (2004).

\bibitem{Bryksin}
P.Keinert, V.V.Bryksin, O. Bleibaum, Phys. Rev. B {\bf 72}, 195311
(2005).

\bibitem{experiment}
G. Engels, J. Lange, Th. Schapers, {\it et al.}, Phys. Rev. B {\bf
55}, 1958R (1997); B. Das, D. C. Miller, S. Datta, {\it et al.},
Phys. Rev. B {\bf 39}, 1411 (1989); J. Luo, H. Munekata, F. F.
Fang {\it et al.}, Phys. Rev. B {\bf 41}, 7685 (1990).

\bibitem{Ganichev04} S.D. Ganichev, V.V. Bel'kov, L.E. Golub, {\it et al.},
Phys. Rev. Lett {\bf 92}, 256601 (2004).

\bibitem{Hall} S. Murakami, N. Nagaosa, and S.C. Zhang, Science {\bf 301},
1348 (2003); J. Wunderlich, B. Kaestner, J. Sinova, {\it et al.},
Phys. Rev. Lett. {\bf 94}, 047204 (2005).

\bibitem{Ganichev03} S.D. Ganichev, V.V. Bel'kov, P. Schneider, {\it et al.},
Phys. Rev. B {\bf 68}, 035319 (2003).

\bibitem{Miller03} J.B. Miller, D.M. Zumb\"uhl, C.M. Marcus, {\it et al.},
Phys. Rev. Lett. {\bf 90}, 076807 (2003).

\bibitem{zitt}
J. Schliemann, D. Loss, R.M. Westervelt, Phys. Rev. Lett. {\bf 94}, 206801 (2005).

\end{thebibliography}
\end{document}